\documentclass[conference]{IEEEtran}
\IEEEoverridecommandlockouts
\usepackage{cite}
\usepackage{amsmath,amssymb,amsfonts}
\usepackage{algorithmic}
\usepackage{graphicx}
\usepackage{textcomp}
\usepackage{xcolor}
\usepackage{verbatim}
\usepackage{colortbl}
\usepackage{booktabs}
\usepackage[hyphens]{url}
\usepackage{hyperref}
\usepackage[hyphenbreaks]{breakurl}

\usepackage{balance}

\def\BibTeX{{\rm B\kern-.05em{\sc i\kern-.025em b}\kern-.08em
    T\kern-.1667em\lower.7ex\hbox{E}\kern-.125emX}}

\usepackage{tikz}
\usepackage{textcomp}

\newcommand\copyrighttext{%
  \footnotesize \textcopyright 2023 IEEE. Personal use of this material is permitted.
  Permission from IEEE must be obtained for all other uses, in any current or future
  media, including reprinting/republishing this material for advertising or promotional
  purposes, creating new collective works, for resale or redistribution to servers or
  lists, or reuse of any copyrighted component of this work in other works.
  \par Accepted to be published in the 32nd International Conference on Computer Communications and Networks (ICCCN 2023), July 24 – 26, 2023, Waikiki Beach, Honolulu, HI, USA.}
\newcommand\copyrightnotice{%
\begin{tikzpicture}[remember picture,overlay]
\node[anchor=south,yshift=10pt] at (current page.south) {\fbox{\parbox{\dimexpr\textwidth-\fboxsep-\fboxrule\relax}{\copyrighttext}}};
\end{tikzpicture}%
}

\begin{document}

\title{Availability Model of a 5G-MEC System\\
\thanks{This work was supported by the Norwegian Research Council through the 5G-MODaNeI project (no. 308909).}
}

\author{\IEEEauthorblockN{Thilina Pathirana}
\IEEEauthorblockA{\textit{Dept of Electrical Eng. and Computer Science} \\
\textit{University of Stavanger}\\
Stavanger, Norway \\
thilina.pathirana@uis.no}
\and

\IEEEauthorblockN{Gianfranco Nencioni}
\IEEEauthorblockA{\textit{Dept of Electrical Eng. and Computer Science} \\
\textit{University of Stavanger}\\
Stavanger, Norway \\
gianfranco.nencioni@uis.no}

}

\maketitle
\copyrightnotice

\begin{abstract}
Multi-access Edge Computing (MEC) is one of the enabling technologies of the fifth generation (5G) of mobile networks. MEC enables services with strict latency requirements by bringing computing capabilities close to the users.
As with any new technology, the dependability of MEC is one of the aspects that need to be carefully studied.
In this paper, we propose a two-level model to compute the availability of a 5G-MEC system.
We then use the model to evaluate the availability of a 5G-MEC system under various configurations.
The results show that having a single redundancy of the 5G-MEC elements leads to an acceptable availability.
To reach a high availability, the software failure intensity of the management elements of 5G and MEC should be reduced.

\end{abstract}


\section{Introduction}

The fifth generation (5G) of mobile networks provides new advanced services, such as Ultra-Reliable Low-Latency Communication (URLLC).
Multi-access Edge Computing (MEC) is one of the technologies that enable such services. MEC brings computational and storage capabilities close to the end user. MEC enables low latency, cloud offloading, and context awareness~\cite{ETSI:5G:MEC}.
URLLC services require not only low latency but also high dependability. Availability is one of the attributes of dependability, and availability refers to the ability of a system to be operational and accessible when needed~\cite{Avizienis:taxonomy,Trivedi:DepSec:2009}. 
Availability is a critical aspect of system design and is measured in terms of probability of a system to be up or in terms of uptime percentage over a given period. For example, a URLLC service requires the availability of 0.99999 or 99.999\%, i.e., the so-called 5-nines availability~\cite{five9s}, which means the system is operational for all but 5.26 minutes per year, while a system with an availability of 0.99 or 99\% could be down for up to 3.65 days per year.
Since MEC will support URLLC in 5G networks, it is crucial that the 5G and MEC will be able to guarantee the required availability. For this reason, a model of the availability of the 5G-MEC system is important to evaluate the ability of the system to meet such strict requirements.

However, the overall availability of the 5G-MEC system as a whole has not been extensively studied yet. While there have been some works on modeling the availability of 5G MEC~\cite{mec_survey}, many of them are focused on specific services or components of the system. For instance, ~\cite{8254591} examines the availability of Virtual Machines (VMs) on the edge and formulates a model for VM and host failures, but its primary focus is on cost optimization rather than the actual availability of the system. Similarly, ~\cite{8647858} introduces an availability model to compute the availability of service deployment in MEC, but it primarily deals with resource allocation problems.
In~\cite{dept_opt}, the authors model the availability of a system composed of an edge server and a cloud server by using a simple continuous-time Markov chain.
In~\cite{cloudlet}, the authors model the availability of a system composed of a cloudlet and a cloud by using a Stochastic Reward Network. The model includes details on the cloudlet/cloud failure modes.
In~\cite{sfc_aging}, the authors model the availability of a service function chain deployed in a MEC system by using a semi-Markov-process-based approach. The authors focus on the aging aspects.
In~\cite{NguyenmedIoT}, the authors model the availability of a cloud-fog-edge system of the Internet of Medical Things by using a three-level model. The model includes the failure modes of the components of each element of the system and also includes security aspects.
All these works do not model the access network and do not consider the management elements. For most of them, MEC is only evaluated together with the cloud to perform task offloading, other works focus on specific scenarios. None of these works model the failures and interactions of all the elements in both 5G and MEC.

This paper fills this gap and the main contribution is twofold: 
\begin{itemize}
    \item We propose a two-level availability model of the 5G-MEC system. 
    
    The model do not assume any specific service, but it is particular important for URLLC services, which have strict availability requirements.
     
    The model is able to represent in a scalable way the interactions between 5G and MEC elements (including the management) and the failure modes of each element.
    \item We perform an evaluation to investigate the impact on the system availability of the redundancy of the 5G and MEC elements and the failure intensities of the management elements. This evaluation will suggest how a 5G/MEC system should be designed in order to increase the availability and match the strict requirement of a URLLC application.
\end{itemize}

The rest of the paper is organized as follows. Section~\ref{sec:model} introduces the proposed availability model of 5G-MEC system. Section~\ref{sec:eval} presents the evaluation of the availability of the 5G-MEC system under various configurations. Finally, Section~\ref{sec:conclude} concludes the paper.


\section{Two-level Availability Model} \label{sec:model}

\begin{figure}[htbp]
    \centering
    \includegraphics[width= \columnwidth]{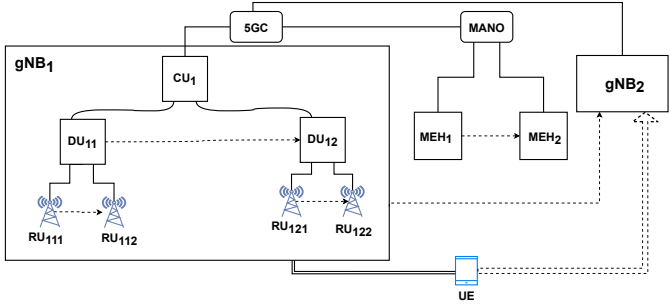}
    \caption{5G-MEC Scenario}
    \label{fig:scenario1}
\end{figure}

In this work, we base on the ETSI architecture for MEC and 5G integration~\cite{ETSI:MEC:031}. In particular, we consider a 5G-MEC scenario as illustrated in Figure~\ref{fig:scenario1}.

The 5G is composed of the 5G Core (5GC) and the 5G Radio Access Network (RAN).
The 5GC has all the essential controlling and management functions of the 5G network including authorizations, authentications, and different policy management. The set of functions composing the 5GC can be deployed in a virtual environment located centrally for a specific operator.
The 5G RAN is composed of 5G base stations, called gNodeB. The User Equipment (UE) connects to one of the gNodeBs by using the 5G New Radio (NR) air interface. A gNodeB will provide user plane and control plane connectivity between the 5GC and the UE hosting a number of protocol layers such as Radio Resource Control (RRC), Packet Data Convergence Control (PDCP), Radio Link Control (RLC), Medium Access layer (MAC) and Physical layer (PHY)~\cite{ETSI:5G:138.300}.
We consider a functional split of the gNodeB: the Central Unit (CU) is a logical centrally-hosted unit handling the RRC, SDAP, and PDCP protocol functions; the Distributed Unit (DU) handles the RLC, MAC, and some PHY protocol functions. The Radio Unit (RU) handles the lower physical-layer connectivity and the radio functionalities. Multiple RUs are connected to one DU and multiple DUs are connected to one CU~\cite{ETSI:5G:138.401}. For the simplicity of the reader, we include a list of all acronyms in Table~\ref{tab0}.

\begin{table}[htbp]
\caption{List of Acronyms}
\begin{center}
\begin{tabular}{ll}
\toprule

5G	&	Fifth Generation of mobile networks\\
5GC	&	5G Core\\
APP	&	Application\\
COTS	&	Commercial-Off-The-Shelf\\
CU	&	Central Unit\\
DU	&	Distributed Unit\\
ETSI &  European Telecommunications Standards Institute\\
FT	&	Fault Tree\\
FW	&	Firmware\\
HW	&	Hardware\\
MANO	&	Management and Orchestration\\
MEC	&	Multi-access Edge Computing \\
MEH	&	MEC Host\\
MEO	&	MEC Orchestrator\\
MEP	&	MEC Platform\\
MEPM	&	MEC Platform Manager\\
NFV	&	Network Function Virtualization\\
NR	&	New Radio\\
OS	&	Operating System\\
RAN	&	Radio Access Network\\
RU	&	Radio Unit\\
SAN	&	Stochastic Activity Network\\
SDN	&	Software-Defined Networking\\
SW	&	Software\\
UE	&	User Equipment\\
URLLC	&	Ultra-Reliable Low-Latency Communication\\
VIM	&	Virtualization Infrastructure Manager\\
VM	&	Virtual Machine\\
\bottomrule
\end{tabular}
\label{tab0}
\end{center}
\end{table}

For the MEC, we consider the ETSI architecture \cite{ETSI:MEC:003}. A MEC system is composed of one or multiple MEC Hosts (MEHs). A MEH has the computational resources and runs one or multiple MEC applications for one or multiple UEs.
The Management and Orchestration (MANO) is composed by the MEC Orchestrator (MEO), which manages and orchestrates at the system level and the MEC Platform Manager (MEPM) and Virtualization Infrastructure Manager (VIM), which manage at host level.

We assume that the UE can connect to all the RUs and use any MEH in order to receive a MEC service. We assume that the interconnections of the 5G and MEC elements are always available, these interconnections can be physical (e.g., 5G NR interface) or logical (i.e., a path in the underlying network connecting the elements). ETSI~\cite{ETSI:MEC:031} suggests different ways of integrating  5G with MEC and deploying the MEHs with the 5G RAN. Since we do not model the interconnections between the 5G and MEC elements, our model is general and does not assume any specific deployment. 

The redundancy consists in duplicating critical elements in a system or components in a element to create a backup in case of failure. There are multiple types of redundancy. 1) \emph{Active-standby redundancy}, where there are two identical elements or systems, one active and the other on standby. The active component handles all the workload, while the standby element is idle but ready to take over if the active one fails. 2) \emph{Active-active redundancy}, where two or more identical elements or systems work in parallel to handle the workload. If one component fails, the others continue to work and take over its workload. 3) \emph{N+k redundancy}, where the system has $k$ (where $k<N$) extra components that are used as a backup for $N$ components. For example, if there are three critical elements, there would be four, with one extra element on standby to take over if any of the three fail~\cite{zhang2010reliability}. The goal of redundancy is to improve a system's overall reliability and availability. By having a backup in place, if one component or system fails, another one is ready to take over, minimizing downtime and disruption to operations. 
The redundancy of the elements and the related components in the 5G-MEC system under investigation will be presented together with the model.

To model the steady-state system availability of this 5G-MEC system, we propose a hierarchical model, which is able to capture the details of the failure process in a scalable manner.

For the top level, we use a Fault Tree (FT) to model the whole 5G-MEC system. The FT model captures the inter-dependencies between the various 5G and MEC elements composing the systems. The FT model shows how the failure of one element impacts the overall system availability.
For the bottom level, we use Stochastic Activity Networks (SANs) to model each element composing the 5G-MEC system. The SAN models capture the multiple failure modes of the 5G and MEC elements.


\subsection{FT Model} 

\begin{figure}[htbp]
    \centering
    \includegraphics[width=\columnwidth]{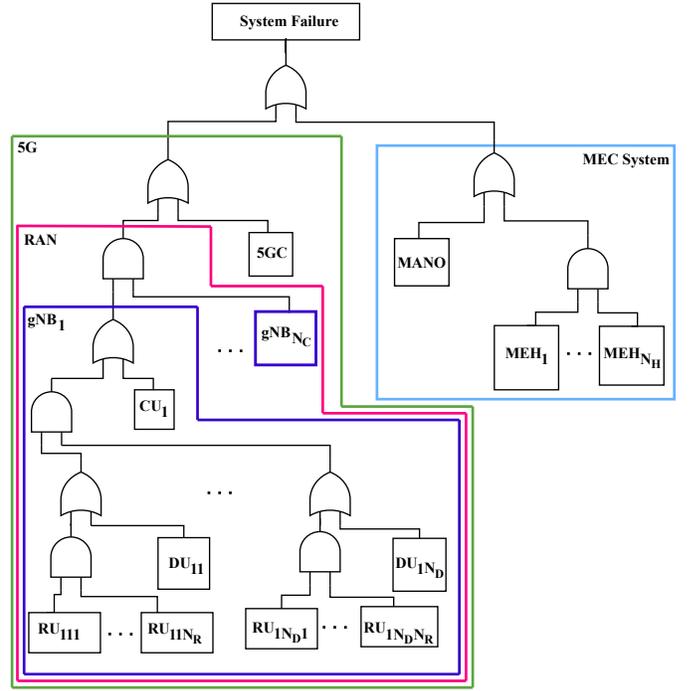}
    \caption{FT model of the 5G-MEC system}
    \label{fig:FTL1}
\end{figure}

To compute the overall 5G-MEC system availability, we propose the FT model depicted in Figure~\ref{fig:FTL1} and the FT explains the main elements of the overall system that may affect a complete system failure. The key elements of the 5G and MEC are only considered for this level and the lowest node in each branch will denote the respective element. The FT model uses the Boolean logic~\cite{ericson1999fault} and with logic gate shows how the various 5G and MEC elements impact the overall system availability. 
We consider an active-active redundancy for RU, DU, CU, and MEH. The UE can connect to any of the $N_H$ MEHs and can be associated to any of the $N_C$ CUs (and therefore gNodeBs). Given a CU, the UE can be associated to any of the $N_D$ DUs. Given a DU, the UE can be associated to any of the $N_R$ RUs.
Only one MEH needs to be available and only one RU (and the related DU and CU) must be available in order for the whole system to be available. 
We instead consider only one 5GC and MANO, they both must be available in order for the whole system to be available.
We consider the MANO as one element as in previous works on Network Function Virtualization (NFV) \cite{tola2019network,tola2019drcn}.

Given the proposed FT, the unavailability of the 5G-MEC system, $U_{Sys}$ can be computed from the unavailability of the various elements (i.e., $U_{5GC}$, $U_{CU}$, $U_{DU}$, $U_{RU}$, $U_{MANO}$, and $U_{MEH}$) as follows:
\begin{equation*}
\resizebox{.97\hsize}{!}{$
U_{Sys} = 1 - \left[\left(1 - U_{RAN}\right)\left(1 - U_{5GC}\right)\left(1 - U_{MANO}\right)\left(1 -  U_{MEH}^{N_M}\right)\right]
$},
\end{equation*}
where
\begin{equation*}
\resizebox{.97\hsize}{!}{$
U_{RAN} = \left[1 - \left(1 - \left(1 - \left(1 - U_{RU}^{N_R}\right)\left(1 - U_{DU}\right)\right)^{N_D}\right)\left(1 - U_{CU}\right)\right]^{N_C}
$}.
\end{equation*}


\subsection{SAN Models of the 5G-MEC Elements}

In the following, we present the SAN models of each 5G-MEC element.
A SAN is a more generalized version of stochastic Petri nets. The use of SAN allows us to model qualities such as repetition, timely responsiveness, and re-usability in the same model~\cite{meyer1985stochastic}. 
Given a similar nature, we model the 5GC and the MANO by using the same model. RU, DU, CU, and MEH have instead a dedicated model. Some models are influenced by works on NFV \cite{tola2019network, tola2019drcn} and Software-Defined Networking (SDN)\cite{nencioni2016dsn,nencioni2017including}. Some concepts are reused within the models to represent similar components.

\subsubsection{RU}
We model RU, but also DU and CU, by considering the OpenRAN specifications along with the implementations discussed by~\cite{telefonica,telecominfra}.
OpenRAN provides an open-source solution for implementing RAN functions on Commercial-Off-The-Shelf (COTS) servers allowing low-cost installations. 

\begin{figure}[htbp]
    \centering
    \includegraphics[width=.8\columnwidth]{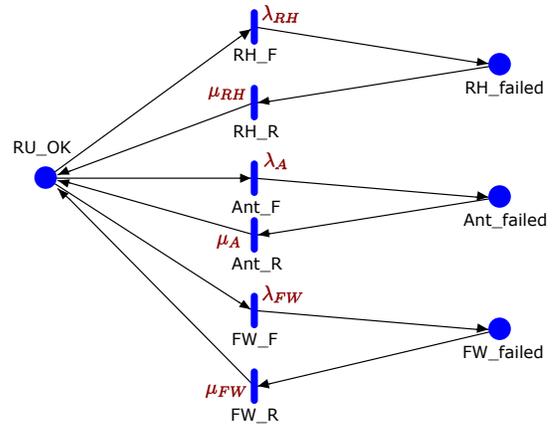}
    \caption{SAN model of a RU}
    \label{fig:SAN:RU}
\end{figure}

Figure~\ref{fig:SAN:RU} shows the proposed SAN model to compute the availability of a RU. The model contains the following places:
\begin{itemize} 
  \item \textit{RU\_OK} represents the fully working state of the RU and is initiated with one token.
  \item \textit{RH\_failed} represents the failure of the hardware (HW) of the RU.
  \item \textit{Ant\_failed} represents the failure of the RU antenna.
  \item \textit{FW\_failed} represents the failure of the RU firmware (FW) that runs on the device on top of the HW.
\end{itemize}

For the SAN model of RU, the following timed activities are used:
\begin{itemize} 
  \item \textit{RH\_F} and \textit{RH\_R} represent the HW failure and recovery events with the rates $\lambda_{RH}$ and $\mu_{RH}$, respectively.
  \item \textit{Ant\_F} and \textit{Ant\_R} represent the failure and recovery events and the for the RU antenna with the rates $\lambda_{A}$ and $\mu_{A}$, respectively.
  \item \textit{FW\_F} and \textit{FW\_R} represent the FW failure and recovery events of the RU with the rates $\lambda_{FW}$ and $\mu_{FW}$, respectively.
\end{itemize}

\subsubsection{DU}

\begin{figure}[htbp]
    \centering
    \includegraphics[width=\columnwidth]{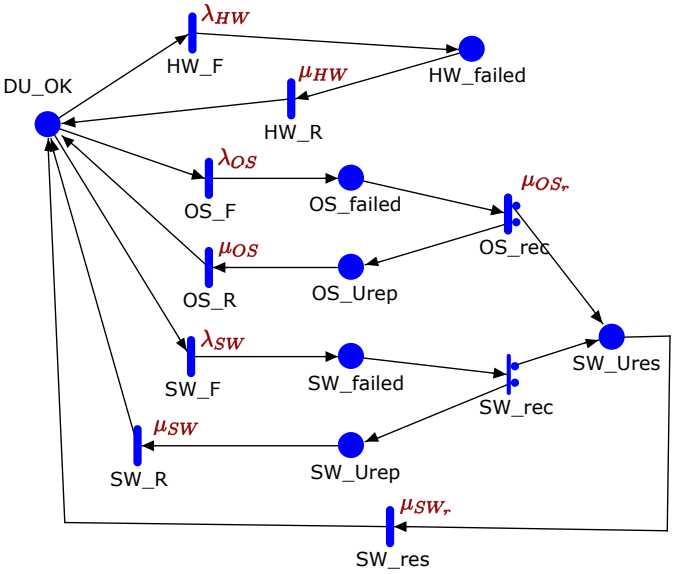}
    \caption{SAN model of a DU}
    \label{fig:SAN:DU}
\end{figure}

Figure~\ref{fig:SAN:DU} represents the SAN model to compute the availability of a DU.
We assume the DU application, which consists of the DU functions, is running on COTS hardware without any redundancy. 
The model contains the following places:
\begin{itemize} 
  \item \textit{DU\_OK} represents the fully working state and is initiated with one token.
  \item \textit{HW\_failed} represents the failure of the DU HW.
  \item \textit{OS\_failed} represents the failure of the generic Operating System (OS) that runs on the DU.
  \item \textit{OS\_Urep} represents the state where the OS undergoes a hard repair process.
  \item \textit{SW\_failed} represents the failure of the software (SW), which implements the DU functions and runs on top of the OS.
  \item \textit{SW\_Urep} represents the state of hard repair for the DU software application. 
  \item \textit{SW\_Ures} represents the state where the DU software application undergoes a restart.
\end{itemize}

The timed activities of the model which connects the above places are described as follows:
\begin{itemize} 
  \item \textit{HW\_F} and \textit{HW\_R} represents the HW failure and recovery events with the rates $\lambda_{HW}$ and $\mu_{HW}$, respectively.
  \item \textit{OS\_F} and \textit{OS\_R} represents the failure and recovery events for the DU OS with the rates $\lambda_{OS}$ and $\mu_{OS}$, respectively.
  \item \textit{OS\_rec} represents the recovery event for the DU OS at the rate $\mu_{{OS}_r}$. It consists of a simple OS reboot and there are two cases, with probability $C_{OS}$ a simple reboot successfully recovers the failure, and with probability, $1-C_{OS}$ the reset is not successful therefore a hard repair is needed. In the first case, the token is fetched from \textit{OS\_failed} to \textit{SW\_Ures}, this is because after the reboot the SW needs to be restarted. In the second case, the token will be moved to \textit{OS\_Urep}.
  \item \textit{SW\_F} and \textit{SW\_R} represents the SW application failure and hard repair events with the rates $\lambda_{SW}$ and $\mu_{SW}$, respectively.
  \item \textit{SW\_rec} represents the instantaneous activity for the recovery event of the DU SW application. This is a two-case activity where with probability $C_{SW}$ an SW restart is enough to recover the failure and the token is fetched from \textit{SW\_failed} to \textit{SW\_Ures}. With probability $1-C_{SW}$, the restart is not enough and the token will be moved to \textit{SW\_Urep} for the hard repair.
  \item \textit{SW\_res} represents the SW restart event with the rate $\mu_{{SW}_r}$.
\end{itemize}

\subsubsection{CU}
\begin{figure}[htbp]
    \centering
    \includegraphics[width=\columnwidth]{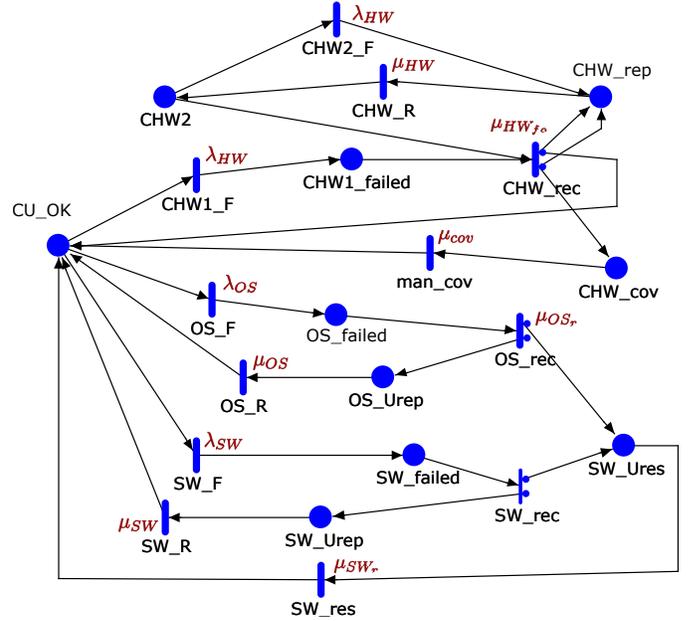}
    \caption{SAN model of a CU}
    \label{fig:SAN:CU}
\end{figure}

We consider an implementation of the CU that is similar to the DU but has a 1+1 active-standby HW redundancy.
Figure~\ref{fig:SAN:CU} shows the SAN model of the CU. The model has places and timed activities that are the same of the DU model. The  different places are the following.
\begin{itemize} 
  \item \textit{CU\_OK} represents the fully working state of CU and is initiated with one token.
  \item \textit{CHW2} represents that the secondary HW is ready to take over in case of HW failure and is initiated with one token.
  \item \textit{CHW1\_failed} represents the failure of the primary HW.
  \item \textit{CHW\_rep} represents the repair state of an HW component.
  \item \textit{CHW\_cov} represents the state where the failover is not successful so manual intervention is needed. 
\end{itemize}

The timed activities of the CU model that are different from the DU model are the following.
\begin{itemize} 
  \item \textit{CHW1\_F} and \textit{CHW2\_F} represents the primary and secondary HW failure events, respectively with the rates $\lambda_{HW}$.
  \item \textit{CHW\_R} represents the HW recovery event with the rate $\mu_{HW}$.
  \item \textit{CHW\_rec} represents the failover event of the CU HW at the rate $\mu_{{HW}_{fo}}$. There are two cases, with probability $C_{HW}$ the failover is successful and a token is fetched from the \textit{CHW1\_failed} to \textit{CHW\_rep}  and another token from \textit{CHW2} to \textit{CU\_OK} implying that the HW redundancy works and the system is back on the normal working state. With probability $1-C_{HW}$, the failover is unsuccessful and one token will be moved to \textit{man\_cov} and the other token to the \textit{CHW\_rep}.
  \item \textit{man\_cov} represents the manual intervention event after all hardware fails with the rate $\mu_{cov}$.

\end{itemize}

\subsubsection{MEH}

We assume that a MEH is a virtualization-capable COTS computing system \cite{lee2019case}. MEHs have a type-II hypervisor system, where the MEC application runs on top of a Virtual Machine (VM)\cite{ETSI:MEC:027}. According to the ETSI MEC architecture \cite{ETSI:MEC:003}, the MEH includes a MEC Platform (MEP), which has management functionalities and can be seen as an SW running on top of another VM.

\begin{figure}[htbp]
    \centering
    \includegraphics[width=\columnwidth]{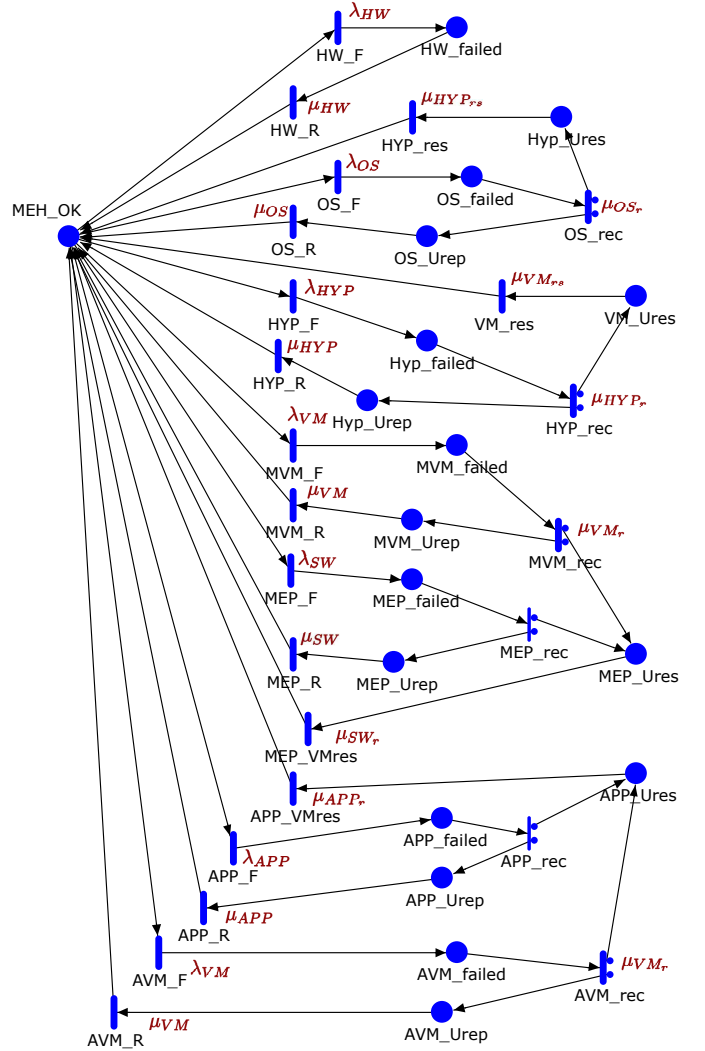}
    \caption{SAN model of a MEH }
    \label{fig:SAN:MEH}
\end{figure}

The SAN model of the MEH is shown in Figure~\ref{fig:SAN:MEH} and the additional places are defined as follows:
\begin{itemize} 
  \item \textit{MEH\_OK} represents the fully working state and is initiated with one token.
  \item \textit{Hyp\_failed} represents the failure of the hypervisor.
  \item \textit{Hyp\_Ures} represents the state where the hypervisor undergoes a soft restart continued with the restart of VMs and applications of MEP/APP.   
  \item \textit{Hyp\_Urep} represents the state of hard repair of the hypervisor.
  \item \textit{VM\_Ures} represents the state where the VM undergoes a soft restart continued with the restart of MEP/APP.
  \item \textit{MVM\_failed} represents the failure of the VM hosting the MEP and \textit{MVM\_Urep} represents the state of hard repair of the MEP-VM.
  \item \textit{MEP\_failed} and \textit{MEP\_Urep} represent the failure and state of hard repair of the MEP software respectively and \textit{MEP\_Ures} represents the state where the MEP undergoes a restart.  
  \item \textit{AVM\_failed} and \textit{AVM\_Urep} represent the failure and the state of hard repair of the VM hosting the MEC application, respectively.
  \item \textit{APP\_failed} and \textit{APP\_Urep} represent the failure and the state of hard repair of the MEC application, respectively. \textit{APP\_Ures} represents the state where the MEC application undergoes a restart.
\end{itemize}

The above places are connected with the following timed activities:
\begin{itemize} 
   \item \textit{HYP\_res} represents the restart event of hypervisor, VMs, and MEP/APP with the rate $\mu_{{HYP}_{rs}}$.
   \item \textit{HYP\_F} and \textit{HYP\_R} represent the hypervisor failure and hard repair events with the rates $\lambda_{HYP}$ and $\mu_{HYP}$, respectively.
   \item \textit{HYP\_rec} represents the recovery event for the hypervisor at the rate $\mu_{{HYP}_r}$. It consists of a simple hypervisor restart and there are two cases, with probability $C_{HYP}$ a simple restart successfully recovers the failure and with probability, $1-C_{HYP}$ the restart is not successful therefore a hard repair is needed. In the first case, the token is fetched from \textit{HYP\_failed} to \textit{VM\_Ures}, this is because after the hypervisor restart the VM needs to be also restarted. In the second case, the token move to \textit{HYP\_Urep} for the hard repair. 
   \item \textit{VM\_res} represents the restart event of VMs and the MEP/APP with the rate $\mu_{{VM}_{rs}}$.
   \item \textit{MVM\_F} and \textit{MVM\_R} represent the failure and hard repair events of the VM hosting MEP software with the rates $\lambda_{VM}$ and $\mu_{VM}$, respectively.
   \item \textit{MVM\_rec} represents the recovery event for failed MEP hosting VM at the rate $\mu_{{VM}_r}$. With probability $C_{VM}$, the MEP VM restart is successful and the token is fetched from \textit{MVM\_failed} to \textit{MEP\_Ures}. With probability, $1-C_{VM}$,  a hard repair is needed and the token will be moved to \textit{MVM\_Urep}.
   \item \textit{MEP\_F} and \textit{MEP\_R} represents the MEP failure and hard repair events with the rates $\lambda_{SW}$ and $\mu_{SW}$, respectively.
   \item \textit{MEP\_rec} represents the recovery event for MEP and it is an instantaneous activity. This is a two-case activity and with the probability $C_{APP}$ the MEP restart is enough and the token is fetched from  \textit{MEP\_failed} to {MEP\_Ures}. With probability, $1-C_{APP}$, a hard repair is needed and the token will be moved to {MEP\_Urep}.
   \item \textit{MEP\_VMres} and \textit{APP\_VMres} represents the MEP VM restart event and the application VM restart with the rates $\mu_{{SW}_{r}}$ and $\mu_{{APP}_{r}}$, respectively.
   \item \textit{APP\_F} and \textit{APP\_R} represents APP failure and hard repair events with the rates $\lambda_{APP}$ and $\mu_{APP}$, respectively.
   \item \textit{APP\_rec} represents the recovery event for the application and it is an instantaneous activity. This is a two-case activity and with the probability, $C_{APP}$ the application restart is enough and the token is fetched from \textit{APP\_failed} to {APP\_Ures}. With probability, $1-C_{APP}$, a hard repair is needed and the token will be moved to {APP\_Urep}. 
   \item \textit{AVM\_F} and \textit{AVM\_R} represents the failure and the hard repair events of the VM hosting the MEC application with the rates $\lambda_{VM}$ and $\mu_{VM}$, respectively.
   \item \textit{AVM\_rec} represents the recovery event for the VM hosting the application at the rate $\mu_{{VM}_r}$. This is a two-case activity and with the probability, $C_{VM}$ the VM restart is enough successful, and the token is fetched from the \textit{AVM\_failed} state to \textit{APP\_Ures} state. With probability, $1-C_{VM}$, a hard repair is needed and the token will be moved to \textit{AVM\_Urep} state.
\end{itemize}

\subsubsection{5GC/MANO} 

\begin{figure}[htbp]
    \centering
    \includegraphics[width=\columnwidth]{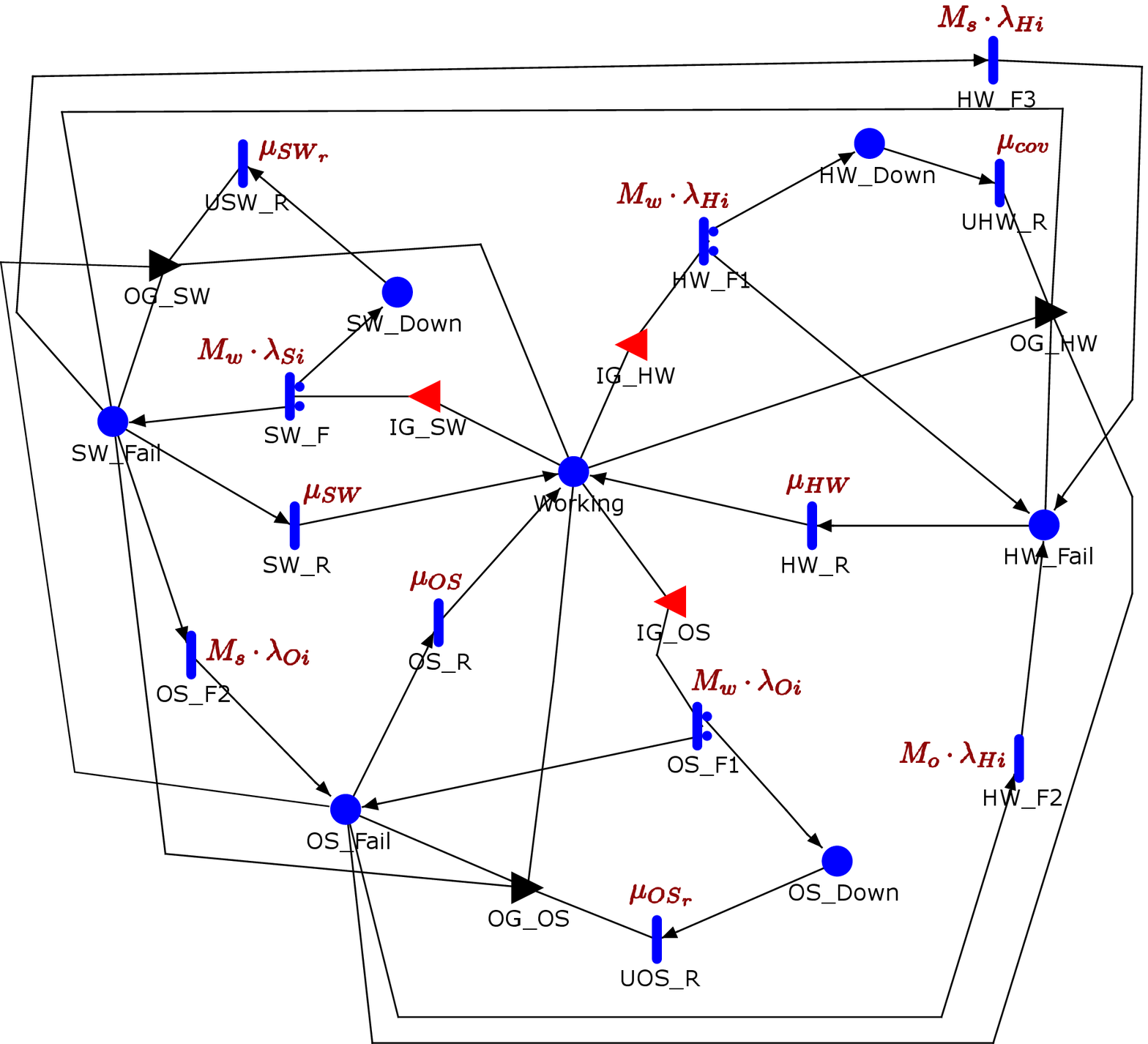}
    \caption{SAN model of the 5GC/MANO }
    \label{fig:SAN:MEO}
\end{figure}

Since 5GC and MANO have similar implementations in a data-center environment and we use the same SAN model, which is shown in Figure~\ref{fig:SAN:MEO}. We assume that the SW implementing the functions of 5GC or MANO is running on top of a generic operating system in a centralized environment~\cite{9670414,7931566}. For this work, the setup is mainly inspired by the Open Source MANO (OSM) which is for the orchestration of NFV~\cite{9324847}. Hereafter for simplicity we refer only to MANO, but the same is valid for the 5GC. 
The data-center environment is modeled as a cluster of $M$ instances and we consider that the system is working if $K\leq M$ instances are working. 
This means that the cluster provides a N+k active-active redundancy, where $N=M-k$.
We also consider the same cluster assumptions as in \cite{tola2019network}. 

The places in the model can be described as follows.
\begin{itemize} 
    \item \textit{Working} represents the number of working instances in the MANO cluster, which is indicated with $M_w$ and is initialized with $M$ tokens.
    \item \textit{HW\_Fail} represents the number of instances where the HW has failed. 
    \item \textit{HW\_Down} is a coverage state and has one token if the HW failure in one instance forces the whole system to
be down.
    \item \textit{OS\_Fail} represents the number of instances where the OS has failed, indicated as $M_{o}$.
    \item \textit{OS\_Down} is a coverage state and has one token if the OS failure in one instance forces the system to
be down.
    \item \textit{SW\_Fail} represents the number of instances where the SW has failed, indicated as $M_{s}$.
    \item \textit{SW\_Down} is a coverage state and has one token if the SW failure in one instance forces the system to
be down.
\end{itemize}

The above places are connected with the following timed activities:
\begin{itemize} 
    \item \textit{HW\_F1} represents the HW failure at a rate of $M_w\cdot\lambda_{Hi}$ where $\lambda_{Hi} = \alpha_H \cdot \lambda_{HW}\cdot M/K$. It has two cases, with the probability $C_{HW}$, the HW failover is successful and a token goes to \textit{HW\_Fail}, otherwise, the system goes down and a token moves to \textit{HW\_Down}.
    \item \textit{HW\_F2} represents the HW failure in instances with a failed OS with a rate of $M_o\cdot\lambda_{Hi}$.
    \item \textit{HW\_F3} represents the HW failure in instances with a failed SW with a rate of $M_s\cdot\lambda_{Hi}$.
    \item \textit{HW\_R} represents the HW recovery with the rate of $\mu_{HW}$.
    \item \textit{UHW\_R} represents the recovery after an unsuccessful HW failover with a rate of $\mu_{cov}$.
    \item \textit{OS\_F1} represents the OS failure at a rate of $M_w\cdot\lambda_{Oi}$ where $\lambda_{Oi}=\alpha_O\cdot\lambda_{OS}\cdot M/K$.
    It has two cases, with the probability $C_{OS}$, the OS failover is successful and a token goes to \textit{OS\_Fail},  otherwise, all the system goes down and a token goes to \textit{OS\_Down}.
    \item \textit{OS\_F2} represents the OS failure in instances with a failed SW with a rate of $M_s\cdot\lambda_{Oi}$. 
    \item \textit{OS\_R} represents the recovery of OS with the rate of $\mu_{OS}$.
    \item \textit{UOS\_R} represents the recovery from the OS crash with a rate of $\mu_{{OS}_r}$.
    \item \textit{SW\_F} represents the SW failure at a rate of $M_w\cdot\lambda_{Si}$ where $\lambda_{Si}=\alpha_S\cdot\lambda_{SW}\cdot M / M_w$, if $M_w\geq K$, or $\lambda_{Si}=\alpha_S\cdot\lambda_{SW}\cdot M$, otherwise. It has two cases, with the probability $C_{SW}$, the SW failover is successful and a token goes to \textit{SW\_Fail}, otherwise, the whole system goes down and a token moves to \textit{SW\_Down}.
    \item \textit{SW\_R} represents the recovery of SW with a rate of $\mu_{SW}$.
    \item \textit{USW\_R} represents the recovery after an unsuccessful SW failover with a rate of $\mu_{{SW}_r}$.
\end{itemize}

The input and output gates in the model are controlling the activities as follows: 
\begin{itemize} 
    \item \textit{IG\_HW}, \textit{IG\_OS}, and \textit{IG\_SW} enable \textit{HW\_F1}, \textit{OS\_F1}, and \textit{SW\_F1}, respectively, only if $M_w>0$ and there are no tokens in \textit{HW\_Down}, \textit{OS\_Down}, \textit{SW\_Down}. They also imply the decrease of $M_w$.  This means that, when the system is crashed, no further failures will happen.
    \item \textit{OG\_HW} increases the number of tokens in \textit{HW\_Fail}, resets the tokens in \textit{OS\_Fail} and \textit{SW\_Fail}, and sets $M_w$ equal to $M$ minus the number of tokens in \textit{HW\_Fail}. This means that, after the system is recovered after a crash, all the failed OS and SW would be recovered.
    \item \textit{OG\_OS} and \textit{OG\_SW} resets the tokens in \textit{OS\_Fail} and \textit{SW\_Fail}, and sets $M_w$ equal to $M$ minus the number of tokens in \textit{HW\_Fail}.
\end{itemize}

The whole MANO system is considered down if $M_w$~$<$~$K$ or there are tokens in \textit{HW\_Down}, in \textit{OS\_Down}, or in \textit{SW\_Down} states. For the sake of the evaluations, we use the multiplicative factors $\alpha_H$, $\alpha_O$, and $\alpha_S$ to study the effects of failure rates of HW, OS, and SW of the MANO, respectively.

Finally, Table~\ref{tb:1} lists all the default values that were used to evaluate all the models and they are based on previous works with similar components~\cite{tola2019network, nencioni2017including}. Some of the values are the same across the models because they represent rates in similar components.

\begin{table}[htbp]
\caption{Default intensity values}
\centering
\begin{tabular}{lll} 
\toprule
Intensity &Value &Description [Mean time to]\\ 
\midrule
$1/\lambda_{RH}$        & 17 years                  & RU HW failure         \\
$1/\mu_{RH}$            & 6 hours                   & RU HW repair/recovery       \\ 
$1/\lambda_{HW}$          & 6 months                  & HW failure                \\ 
$1/\mu_{cov}$           & 30 minutes                & manual coverage               \\ 
$1/\mu_{HW}$            & 2 hours                   & HW repair                \\ 
$1/\mu_{HW_{fo}}$       & 3 minutes                 & HW failover             \\ 
$1/\lambda_A$           & 104 months                & antenna failure          \\ 
$1/\mu_{A}$             & 6 hours                   & antenna repair/recovery         \\
$1/\lambda_{FW}$        & 75 days                   & FW failure            \\ 
$1/\mu_{FW}$            & 65 minutes                & FW repair/recovery          \\ 
$1/\lambda_{OS}$        & 2 months                  & OS failure           \\ 
$1/\mu_{OS}$            & 1 hour                    & OS repair          \\ 
$1/\mu_{OS_{r}}$        & 1 minute                  & OS reboot          \\ 
$1/\mu_{HYP_{rs}}$      & 2.5 minutes               & restart of hypervisor and VMs          \\
$1/\lambda_{HYP}$       & 4 months                  & hypervisor failure              \\ 
$1/\mu_{HYP}$           & 1 hour                    & hypervisor repair            \\ 
$1/\mu_{HYP_{r}}$       & 1 minute                  & hypervisor restart             \\ 
$1/\mu_{VM_{rs}}$       & 1.5 minute                 & restart of VMs          \\
$1/\lambda_{VM}$        & 3 months                  & VM failure  \\ 
$1/\mu_{VM}$            & 1 hour                    & VM repair  \\ 
$1/\mu_{VM_{r}}$        & 1 minutes                 & VM reboot              \\ 
$1/\lambda_{APP}$       & 2 weeks                   & application failure             \\ 
$1/\mu_{APP}$           & 30 minutes                & application repair             \\ 
$1/\mu_{APP_{r}}$       & 15 seconds                & application software restart              \\ 
$1/\lambda_{SW}$        & 1 month                   & SW failure           \\ 
$1/\mu_{SW}$            & 30 minutes                & SW repair           \\ 
$1/\mu_{SW_{r}}$        & 30 seconds                & SW restart        \\ 
$C_{HW}$                & 0.97                      & coverage factor for HW failover        \\
$C_{OS}$                & 0.9                       & coverage factor for OS reboot/failover              \\ 
$C_{HYP}$               & 0.9                       & coverage factor for hypervisor restart              \\ 
$C_{SW}$                & 0.85                      & coverage factor for SW restart/failover             \\ 
$C_{VM}$                & 0.9                       & coverage factor for VM reboot            \\
$C_{APP}$               & 0.8                      & coverage factor for APP restart              \\ 
$(M,K)$                 & (10,9)                   & cluster settings             \\
\bottomrule
\end{tabular}
    
    \label{tb:1}
\end{table}


\section{Evaluation} \label{sec:eval}

In this section, we numerically evaluate the unavailability of the 5G-MEC system under various configurations. The SAN models have been implemented by using M\"{o}bius~\cite{clark2001mobius}. 
Note that we could not compare it with other works since there are not other works that model the whole 5G-MEC system considering the various elements and their components.

First, we evaluate the impact on the 5G-MEC system availability of possible redundancy setups of the 5G-MEC system. 
As presented in the previous section and depicted in Figure~\ref{fig:scenario1}, a 5G-MEC system can have multiple redundancy configurations. A UE can connect to multiple gNodeBs, which can have multiple DUs and RUs, and can connect to multiple MEHs. 
Moreover, the cluster of the 5GC/MANO can have more instances than the ones required.
To understand the effect of redundancy on the system unavailability, we vary the redundancy configuration.
Finally, since the MANO and 5GC are the single points of failure of the system, we evaluate the impact of HW, OS, and SW failure rates by varying the multiplicative factors $\alpha_H$, $\alpha_O$, and $\alpha_S$.

\begin{table}[htbp]
\caption{System unavailability varying the 5G-RAN and MEH redundancy}
\centering
\begin{tabular}{lllll} 
\toprule
 $N_C$ & $N_D$ & $N_R$ & $N_H$ & System Unavailability ($\times10^{-4}$) \\ 
\midrule
 \textbf{1} & 1 & 1 & 1 & 13.222\\
 \textbf{2} & 1 & 1 & 1 & 2.593\\
 \textbf{3} & 1 & 1 & 1 & 2.582\\
 \textbf{1} & 2 & 2 & 2 & 3.277\\
 \textbf{2} & 2 & 2 & 2 & 1.096\\
 \textbf{3} & 2 & 2 & 2 & 1.095\\
 \textbf{1} & 3 & 3 & 3 & 3.276\\
 \textbf{2} & 3 & 3 & 3 & 1.095\\
 \textbf{3} & 3 & 3 & 3 & 1.095\\
\midrule
 1 & \textbf{1} & 1 & 1 & 13.222\\ 
 1 & \textbf{2} & 1 & 1 & 4.771\\
 1 & \textbf{3} & 1 & 1 & 4.763\\
 2 & \textbf{1} & 2 & 2 & 1.096\\
 2 & \textbf{2} & 2 & 2 & 1.096\\
 2 & \textbf{3} & 2 & 2 & 1.096\\
 3 & \textbf{1} & 3 & 3 & 1.095\\
 3 & \textbf{2} & 3 & 3 & 1.095\\
 3 & \textbf{3} & 3 & 3 & 1.095\\
\midrule
 1 &	1 &	\textbf{1} &	1 &	13.222\\
 1 &	1 &	\textbf{2} &	1 &	6.041\\
 1 &	1 &	\textbf{3} &	1 &	6.036\\
 2 &	2 &	\textbf{1} &	2 &	1.096\\
 2 &	2 &	\textbf{2} &	2 &	1.096\\
 2 &	2 &	\textbf{3} &	2 &	1.096\\
 3 &	3 &	\textbf{1} &	3 &	1.095\\
 3 &	3 &	\textbf{2} &	3 &	1.095\\
 3 &	3 &	\textbf{3} &	3 &	1.095\\
\midrule
 1 &	1 &	1 &	\textbf{1} &	13.222\\
 1 &	1 &	1 &	\textbf{2} &	1.174\\
 1 &	1 &	1 &	\textbf{3} &	1.174\\
 2 &	2 &	2 &	\textbf{1} &	2.583\\
 2 &	2 &	2 &	\textbf{2} &	1.096\\
 2 &	2 &	2 &	\textbf{3} &	1.095\\
3 &	3 &	3 &	\textbf{1} &	2.583\\
 3 &	3 &	3 &	\textbf{2} &	1.095\\
 3 &	3 &	3 &	\textbf{3} &	1.095\\
\bottomrule
\end{tabular}
    
    \label{tb:eval:CU}
\end{table}

Table~\ref{tb:eval:CU} shows the system unavailability under various redundancy configurations for 5G-RAN and MEHs. The table shows that the configuration without any redundancy and the configurations with redundancy on multiple elements have a difference of one order of magnitude.
The element that has the highest impact is the CU/gNodeB. When the UE is able to connect to two gNodeBs, the system availability has a reduction of almost one order of magnitude, regardless of whether it is composed of only one DU and one RU and there is only one MEH. The element that has the least impact is the RU, its redundancy only halves the unavailability with no redundancy. DUs and MEHs have a similar impact. For all the elements, one redundant element is enough. For this reason, in the following we consider $N_C=N_D=N_R=N_H=2$.

\begin{figure}[htbp]
    \centering
    \includegraphics[width=\columnwidth]{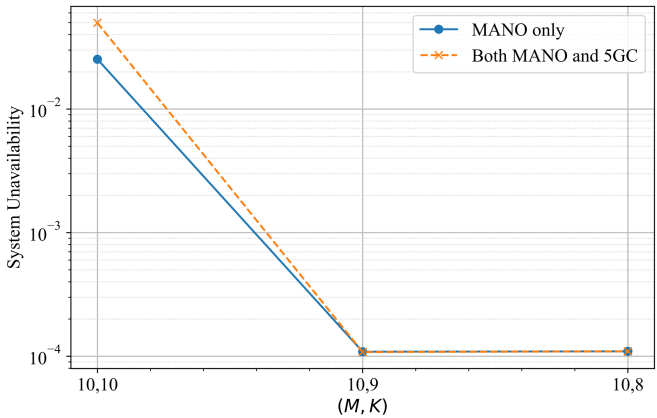}
    \caption{System unavailability varying the cluster redundancy}
   \label{fig:eval:nk10}
\end{figure}

Since the redundancy of 5G-RAN and MEH was evaluated, the next objective was to find how the redundancy of 5GC and MANO affects the total unavailability. Based on the assumptions for the scenario, both 5GC and MANO are hosted in a data center environment and can be seen as a cluster with multiple instances. According to the SAN model of MANO/5GC, at least $K$ out of $M$ working instances should be present to carry out the operations. 
We have varied the $(M,K)$ values of the 5GC/MANO cluster as shown in Figure~\ref{fig:eval:nk10}. The results show an unavailability reduction of two orders of magnitude between $(10,10)$ and $(10,9)$. At $(10,9)$ and $(10,8)$, the unavailability has similar values. Therefore, we must have redundancy in the 5GC/MANO clusters, the redundancy of one instance is enough. This outcome is valid when the redundancy of both MANO and 5GC is varied, but also when the single redundancy of MANO or 5GC is changed.

\begin{figure}[htbp]
    \centering
    \includegraphics[width=\columnwidth]{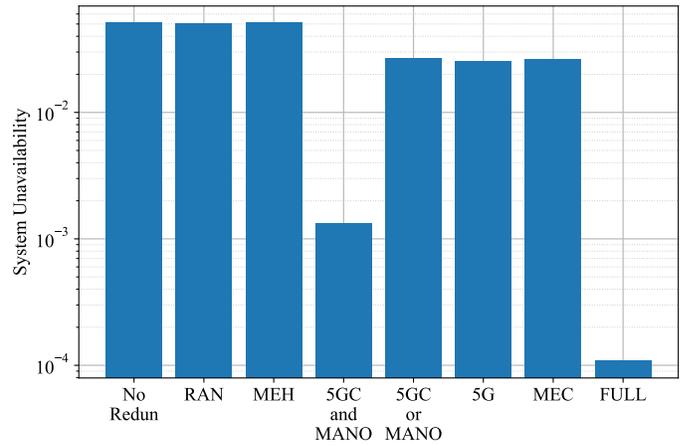}
    \caption{System unavailability for redundancy configurations}
    \label{fig:eval:red}
\end{figure}

Figure~\ref{fig:eval:red} summarizes how the unavailability behaves while varying the different redundancy configurations for the various elements: no-redundancy, partial-redundancy, and full-redundancy. The configurations are set up as follows:
\begin{itemize}
    \item No Redun: $N_C=N_D=N_R=N_H=1$ and $(M,K)=(10,10)$ for both MANO and 5GC,
    \item RAN: $N_C=N_D=N_R=2$, $N_H=1$, and $(M,K)=(10,10)$ for both MANO and 5GC,
    \item MEH: $N_C=N_D=N_R=1$, $N_H=2$, and $(M,K)=(10,10)$ for both MANO and 5GC,
    \item 5GC and MANO: $N_C=N_D=N_R=N_H=1$ and $(M,K)=(10,9)$ for both MANO and 5GC,
    \item 5GC or MANO: $N_C=N_D=N_R=N_H=1$, $(M,K)=(10,9)$ for one of MANO or 5GC and $(M,K)=(10,10)$ for the other,
    \item 5G: $N_C=N_D=N_R=$ and $(M,K)=(10,10)$ for both MANO and 5GC,
    \item MEC: $N_C=N_D=N_R=1$ and $(M,K)=(10,10)$ for both MANO and 5GC, and
    \item FULL: $N_C=N_D=N_R=1$ and $(M,K)=(10,10)$ for both MANO and 5GC.
\end{itemize}

The results show that the redundancy of 5G-RAN or MEH does not significantly affect the unavailability compared to the no-redundancy setup. But when the 5G or MEC is considered, there was a significant change similar to the change that happened when either 5GC or MANO was redundant. The full redundant setup obviously had the lowest unavailability as all the elements are able to failover. 

\begin{figure}[htbp]
    \centering
    \includegraphics[width=\columnwidth]{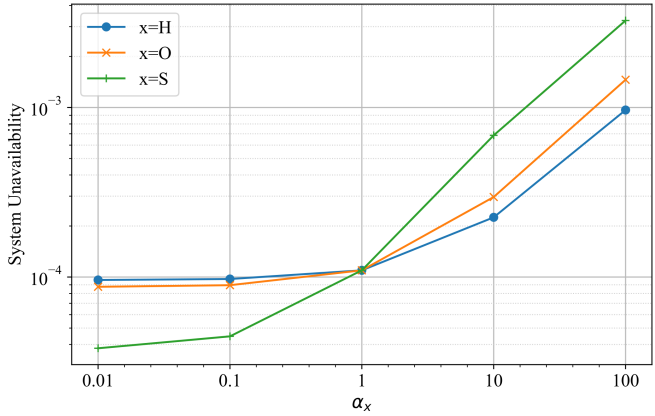}
    \caption{System unavailability varying both 5GC and MANO failure intensities}
    \label{fig:eval:alphaMEO5GC}
\end{figure}

\begin{figure}[htbp]
    \centering
    \includegraphics[width=\columnwidth]{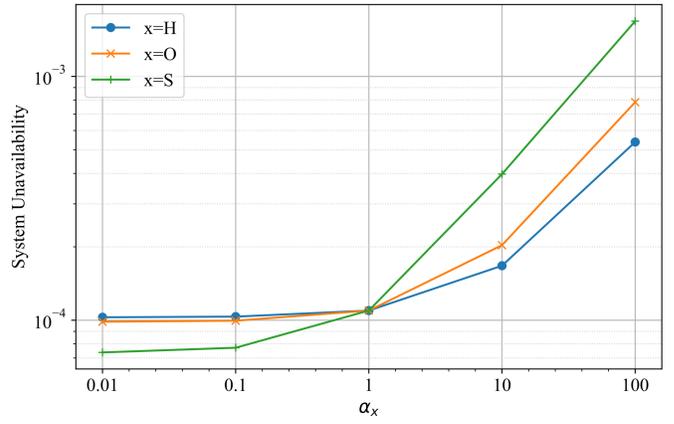}
    \caption{System unavailability varying the single 5GC/MANO failure intensity }
    \label{fig:eval:alphaMEO}
\end{figure}

Therefore, as expected, 5GC and MANO, being single points of failure, have a high impact on the availability of the whole system.
Figure~\ref{fig:eval:alphaMEO5GC} shows the system unavailability when the failure intensities are varied for both 5GC and MANO.
We have varied the failure intensities of HW, OS, and SW by varying the multiplicative factors $\alpha_H$, $\alpha_O$, and $\alpha_S$, respectively.
The figure highlights how an increase in the failure intensities would bring a significant increase in the unavailability, but only the decrease in the SW failure intensity would bring a significant decrease in the unavailability.
We have performed a similar study by varying the $\alpha_x$ factors of the single 5GC/MANO. Figure~\ref{fig:eval:alphaMEO} shows a similar behaviour of the previous figure, but a lower variation when the SW failure intensity is reduced.
In conclusion, both figures suggest that the development of the SW component of MANO and 5GC must consider dependability strategies to reduce the unavailability.

\section{Conclusion} \label{sec:conclude}

In this paper, we have proposed a two-level model to evaluate the availability of a 5G-MEC system.
The model is composed of an FT, which captures the contribution of each 5G and MEC element to the overall availability of the system, and the SAN model of each 5G and MEC element, which captures the failure modes of each element.
Given the proposed model, we have evaluated the impact of the element redundancy on the system availability and the failure intensities of 5GC and MANO.
The results show that each element should be redundant, but a single redundancy is enough. The most critical redundancies are of the RUs/gNodeBs and the clusters of 5GC and MANO. 
In 5GC and MANO, the SW failure intensity is the one that needs to be improved in order to reach the 5-nines availability that is the target of the URRLC services. To reduce the SW failure intensity, several strategies can be implemented, especially during the development, for example, eliminate the single point of failures, enable a reliable crossover, and improve the failure detectability. Overall, a combination of proactive design, monitoring, and planning can help achieve high availability of the software.


\balance
\bibliographystyle{IEEEtran}

\bibliography{bib/IEEEabrv,bib/References}

\begin{thebibliography}{10}
\providecommand{\url}[1]{#1}
\csname url@samestyle\endcsname
\providecommand{\newblock}{\relax}
\providecommand{\bibinfo}[2]{#2}
\providecommand{\BIBentrySTDinterwordspacing}{\spaceskip=0pt\relax}
\providecommand{\BIBentryALTinterwordstretchfactor}{4}
\providecommand{\BIBentryALTinterwordspacing}{\spaceskip=\fontdimen2\font plus
\BIBentryALTinterwordstretchfactor\fontdimen3\font minus
  \fontdimen4\font\relax}
\providecommand{\BIBforeignlanguage}[2]{{%
\expandafter\ifx\csname l@#1\endcsname\relax
\typeout{** WARNING: IEEEtran.bst: No hyphenation pattern has been}%
\typeout{** loaded for the language `#1'. Using the pattern for}%
\typeout{** the default language instead.}%
\else
\language=\csname l@#1\endcsname
\fi
#2}}
\providecommand{\BIBdecl}{\relax}
\BIBdecl

\bibitem{ETSI:5G:MEC}
ETSI, ``{ETSI White Paper No. 28 MEC in 5G networks},'' 06 2018.

\bibitem{Avizienis:taxonomy}
A.~Avizienis, J.-C. Laprie, B.~Randell, and C.~Landwehr, ``Basic concepts and
  taxonomy of dependable and secure computing,'' \emph{IEEE Transactions on
  Dependable and Secure Computing}, vol.~1, no.~1, pp. 11--33, 2004.

\bibitem{Trivedi:DepSec:2009}
K.~S. Trivedi, D.~S. Kim, A.~Roy, and D.~Medhi, ``Dependability and security
  models,'' in \emph{2009 7th International Workshop on Design of Reliable
  Communication Networks}, 2009, pp. 11--20.

\bibitem{five9s}
P.~Popovski, ``Ultra-reliable communication in 5g wireless systems,'' in
  \emph{1st International Conference on 5G for Ubiquitous Connectivity}.\hskip
  1em plus 0.5em minus 0.4em\relax IEEE, 2014, pp. 146--151.

\bibitem{mec_survey}
P.~Maciel, J.~Dantas, C.~Melo, P.~Pereira, F.~Oliveira, J.~Araujo, and
  R.~Matos, ``A survey on reliability and availability modeling of edge, fog,
  and cloud computing,'' \emph{Journal of Reliable Intelligent Environments},
  vol.~8, no.~3, pp. 227--245, 2021.

\bibitem{8254591}
H.~Zhu and C.~Huang, ``Availability-aware mobile edge application placement in
  5g networks,'' in \emph{GLOBECOM 2017 - 2017 IEEE Global Communications
  Conference}, 2017, pp. 1--6.

\bibitem{8647858}
L.~Yala, P.~A. Frangoudis, and A.~Ksentini, ``Latency and availability driven
  vnf placement in a mec-nfv environment,'' in \emph{2018 IEEE Global
  Communications Conference (GLOBECOM)}, 2018, pp. 1--7.

\bibitem{dept_opt}
J.~Liang, B.~Ma, S.~Ali, and J.~Huang, ``Model-based evaluation and
  optimization of dependability for edge computing systems,'' in
  \emph{Collaborative Computing: Networking, Applications and Worksharing},
  H.~Gao and X.~Wang, Eds.\hskip 1em plus 0.5em minus 0.4em\relax Cham:
  Springer International Publishing, 2021, pp. 728--747.

\bibitem{cloudlet}
\BIBentryALTinterwordspacing
H.~Raei and N.~Yazdani, ``Performability analysis of cloudlet in mobile cloud
  computing,'' \emph{Information Sciences}, vol. 388-389, pp. 99--117, 2017.
  [Online]. Available:
  \url{https://www.sciencedirect.com/science/article/pii/S0020025517301330}
\BIBentrySTDinterwordspacing

\bibitem{sfc_aging}
J.~Bai, X.~Chang, F.~Machida, L.~Jiang, Z.~Han, and K.~S. Trivedi, ``Impact of
  service function aging on the dependability for mec service function chain,''
  \emph{IEEE Transactions on Dependable and Secure Computing}, pp. 1--1, 2022.

\bibitem{NguyenmedIoT}
T.~A. Nguyen, D.~Min, E.~Choi, and J.-W. Lee, ``Dependability and security
  quantification of an internet of medical things infrastructure based on
  cloud-fog-edge continuum for healthcare monitoring using hierarchical
  models,'' \emph{IEEE Internet of Things Journal}, vol.~8, no.~21, pp.
  15\,704--15\,748, 2021.

\bibitem{ETSI:MEC:031}
ETSI, ``{GR MEC 031 - V2.1.1 - Multi-access Edge Computing (MEC) MEC 5G
  Integration},'' 10 2020.

\bibitem{ETSI:5G:138.300}
------, ``{ETSI TS 138 300 V16.2.0 - 5G; NR; NR and NG-RAN Overall description;
  Stage-2 (3GPP TS 38.300 version 16.4.0 Release 16)},'' 01 2021.

\bibitem{ETSI:5G:138.401}
------, ``{TS 138 401 - V16.3.0 - 5G; NG-RAN; Architecture description (3GPP TS
  38.401 version 16.3.0 Release 16)},'' 11 2020.

\bibitem{ETSI:MEC:003}
------, ``{GS MEC 003 - V3.1.1 - Multi-access Edge Computing (MEC); Framework
  and Reference Architecture},'' 03 2022.

\bibitem{zhang2010reliability}
X.~Zhang, H.~Pham, and C.~R. Johnson, ``Reliability models for systems with
  internal and external redundancy,'' \emph{International Journal of System
  Assurance Engineering and Management}, vol.~1, pp. 362--369, 2010.

\bibitem{ericson1999fault}
C.~Ericson, ``Fault tree analysis--a history from the proceeding of the 17th
  international system safety conference,'' 1999.

\bibitem{tola2019network}
B.~Tola, G.~Nencioni, and B.~E. Helvik, ``Network-aware availability modeling
  of an end-to-end nfv-enabled service,'' \emph{IEEE Transactions on Network
  and Service Management}, vol.~16, no.~4, pp. 1389--1403, 2019.

\bibitem{tola2019drcn}
B.~Tola, G.~Nencioni, B.~E. Helvik, and Y.~Jiang, ``Modeling and evaluating
  nfv-enabled network services under different availability modes,'' in
  \emph{2019 15th International Conference on the Design of Reliable
  Communication Networks (DRCN)}, 2019, pp. 1--5.

\bibitem{meyer1985stochastic}
J.~F. Meyer, A.~Movaghar, and W.~H. Sanders, ``Stochastic activity networks:
  Structure, behavior, and application.'' in \emph{Unknown Host Publication
  Title}.\hskip 1em plus 0.5em minus 0.4em\relax IEEE, 1985, pp. 106--115.

\bibitem{nencioni2016dsn}
G.~Nencioni, B.~E. Helvik, A.~J. Gonzalez, P.~E. Heegaard, and A.~Kamisinski,
  ``Availability modelling of software-defined backbone networks,'' in
  \emph{2016 46th Annual IEEE/IFIP International Conference on Dependable
  Systems and Networks Workshop (DSN-W)}, 2016, pp. 105--112.

\bibitem{nencioni2017including}
G.~Nencioni, B.~E. Helvik, and P.~E. Heegaard, ``Including failure correlation
  in availability modeling of a software-defined backbone network,'' \emph{IEEE
  Transactions on Network and Service Management}, vol.~14, no.~4, pp.
  1032--1045, 2017.

\bibitem{telefonica}
\BIBentryALTinterwordspacing
F.~Lorca, E.~Serna, M.~Aparacio, A.~Chaissagne, and J.~Espl{\'a}, ``{Telefonica
  views on the design, architecture, and technology of 4G/5G Open RAN networks
  - Open RAN white paper},'' Jan. 2021. [Online]. Available:
  \url{https://www.telefonica.com/es/wp-content/uploads/sites/4/2021/08/Whitepaper-OpenRAN-Telefonica.pdf}
\BIBentrySTDinterwordspacing

\bibitem{telecominfra}
\BIBentryALTinterwordspacing
{Telecom Infra Project Inc.}, ``{OpenRAN 5G NR Base Station Platform
  Requirements Document},'' 2020. [Online]. Available:
  \url{https://telecominfraproject.com/openran/}
\BIBentrySTDinterwordspacing

\bibitem{lee2019case}
D.~Lee and W.~Nam, ``{Case Study of Scaled Up SKT 5G MEC Reference
  Architecture},'' \emph{Intel Builders White Paper}, 2019.

\bibitem{ETSI:MEC:027}
ETSI, ``{GR MEC 027 - V2.1.1 - Multi-access Edge Computing (MEC); Study on MEC
  support for alternative virtualization technologies},'' 11 2019.

\bibitem{9670414}
F.~J. D.~S. Neto, E.~Amatucci, N.~A. Nassif, and P.~A.~M. Farias, ``{Analysis
  for Comparison of Framework for 5G Core Implementation},'' in \emph{2021
  International Conference on Information Science and Communications
  Technologies (ICISCT)}.\hskip 1em plus 0.5em minus 0.4em\relax IEEE, 2021,
  pp. 1--5.

\bibitem{7931566}
T.~Taleb, K.~Samdanis, B.~Mada, H.~Flinck, S.~Dutta, and D.~Sabella, ``On
  multi-access edge computing: A survey of the emerging 5g network edge cloud
  architecture and orchestration,'' \emph{IEEE Communications Surveys \&
  Tutorials}, vol.~19, no.~3, pp. 1657--1681, 2017.

\bibitem{9324847}
M.-I. Csoma, B.~Koné, R.~Botez, I.-A. Ivanciu, A.~Kora, and V.~Dobrota,
  ``Management and orchestration for network function virtualization: An open
  source mano approach,'' in \emph{2020 19th RoEduNet Conference: Networking in
  Education and Research (RoEduNet)}, 2020, pp. 1--6.

\bibitem{clark2001mobius}
G.~Clark, T.~Courtney, D.~Daly, D.~Deavours, S.~Derisavi, J.~M. Doyle, W.~H.
  Sanders, and P.~Webster, ``The mobius modeling tool,'' in \emph{Proceedings
  9th International Workshop on Petri Nets and Performance Models}.\hskip 1em
  plus 0.5em minus 0.4em\relax IEEE, 2001, pp. 241--250.

\end{thebibliography}

\end{document}